\documentclass[aip,cha,reprint,showpacs]{revtex4-1}
\usepackage{graphicx,amsfonts,amsmath,amssymb,color}
\usepackage[english]{babel}

\newcommand{\p}[2]{\frac{\partial\, #1}{\partial\, #2}\,}
\newcommand{\e}{\varepsilon}

\newcommand{\Z}{\mathbb{Z}}

\newcommand{\V}[1]{\mathbf{#1}}
\newcommand{\intl}[2]{\int\limits_{#1}^{#2}}
\newcommand{\bracket}[1]{\left(#1\right)}

\newcommand{\eq}[1]{$\mathrm{Eq.}$~\eqref{#1}}
\newcommand{\figref}[1]{$\mathrm{Fig.}$~\ref{#1}}
\newcommand{\secref}[1]{$\mathrm{Sec.}$~\ref{#1}}

\newcommand{\av}[1]{\left\langle #1 \right\rangle}

\begin{document}

\title{Biased Brownian motion in extreme corrugated tubes}

\author{S. Martens}
\email{steffen.martens@physik.hu-berlin.de}
\affiliation{Department of Physics, Humboldt-Universit\"{a}t zu Berlin, Newtonstr. 15, 12489 Berlin, Germany}
\author{G. Schmid} 
\affiliation{Department of Physics, Universit\"{a}t Augsburg, Universit\"{a}tsstr. 1, 86135 Augsburg, Germany}
\author{L. Schimansky-Geier}
\affiliation{Department of Physics, Humboldt-Universit\"{a}t zu Berlin, Newtonstr. 15, 12489 Berlin, Germany}
\author{P. H\"anggi}
\affiliation{Department of Physics, Universit\"{a}t Augsburg, Universit\"{a}tsstr. 1, 86135 Augsburg, Germany}

\begin{abstract}
\noindent Biased Brownian motion of point-size particles in a three-dimensional tube with smoothly varying cross-section is investigated. In the fashion of our recent work~\cite{Martens2011} we employ an asymptotic analysis to the stationary probability density in a geometric parameter of the tube geometry. We demonstrate that the leading order term is equivalent to the Fick-Jacobs approximation. Expression for the higher order
corrections to the probability density are derived. Using this expansion orders we obtain that in the diffusion dominated regime the average particle current equals the zeroth-order Fick-Jacobs result corrected by a factor including the corrugation of the tube geometry. In particular we demonstrate that this estimate is more accurate for extreme corrugated geometries compared to the common applied method using the spatially dependent diffusion coefficient $D(x,f)$. 
The analytic findings are corroborated with the finite element calculation of a sinusoidal-shaped tube.
\end{abstract}

\maketitle

\begin{quotation}
  Particle transport in micro- and nanostructured channel structures
  exhibits peculiar characteristics which differs from other transport
  phenomena occurring for energetic systems. The theoretical modelling
  involves Fokker-Planck type dynamics in three dimensions which
  cannot be solved for arbitrary boundary conditions imposed by the
  geometrical restrictions. Recently, much effort is drawn on a
  reduction of the complexity of the problem resulting in the
  so-called Fick-Jacobs approximation in which (infinitely) fast
  equilibration in certain spatial directions is assumed. Within the present
  manuscript we derive a reduction method which (i) corresponds in zeroth order in the
  expansion parameter, which describes the corrugation of the tube wall,
  to the celebrated Fick-Jacobs result and (ii) extends the
  validity of the Fick-Jacobs approximation towards extreme corrugated
  tube structures.        
\end{quotation}

\section{Introduction}

The transport of large molecules and small particles that are
geometrically confined within zeolites \cite{Keil2000,Beerdsen2005,Beerdsen2006}, biological\cite{Hille} as well as designed nanopores\cite{Kettner2000,Berezhkovskii2007,Pedone2010}, channels or other
quasi-one-dimensional systems attracted attention in the last
decade. This activity stems from the  profitableness for shape and
size selective catalysis \cite{Cheng2008,riefler2010}, particle separation and the dynamical
characterization of polymers during their translocation
\cite{Muthukumar2001,Matysiak2006,Cacciuto2006,Dekker2007,Hanggi2009}. In
particular, the latter theme which aims at the experimental
determination of the structural properties and the amino acid
sequence in DNA or RNA when they pass through narrow openings or the
so-called bottlenecks, comprises challenges for technical
developments of nanoscaled channel structures
\cite{Dekker2007,Keyser2006,Howorka2009,Hanggi2009}.

Along with the progress of the experimental techniques the problem
of particle transport through corrugated channel structures
containing narrow openings and bottlenecks has give rise to recent
theoretical activities to study diffusion dynamics occurring in such
geometries \cite{Burada2009_CPC}. Previous studies by Jacobs
\cite{Jacobs} and Zwanzig \cite{Zwanzig1992} ignited a revival of
doing research in this topic. The so-called {\it Fick-Jacobs
approach} \cite{Jacobs, Zwanzig1992,Burada2009_CPC}, accounts for the elimination
of  transverse stochastic degrees of freedom by assuming a (infinitely) fast
equilibration in those transverse directions. The theme found its application for biased particle transport through periodic $3$D planar channel structures \cite{Reguera2006,Burada2007, Burada2008,
Kalinay2006,Drazer2010,Martens2011} as well as for tubes \cite{Ai2006,Berezhkovskii2007,Dagdug2010,Dagdug2011} exhibiting smoothly varying side-walls.

Beyond the Fick-Jacobs (FJ) approach, which is suitably applied to
channel geometries with smoothly varying side-walls, there exist yet
other methods for describing the transport through varying channel
structures like cylindrical septate channels
\cite{Marchesoni2010,Borromeo2010,Hanggi2010,Makhnovskii2011}, tubes formed by
spherical compartments \cite{Berezhkovskii2010a, Berezhkovskii2010b} 
or channels containing  abrupt changes of cross diameters
\cite{Kalinay2010,Makhnovskii2010}.

In a recent work\cite{Martens2011} we have provided a systematic treatment by
using a series expansion of the stationary probability density in terms of a geometric parameter which
specifies the channel corrugation for {\it biased} particle
transport proceeding along a {\itshape planar three-dimensional
  channel} exhibiting periodically varying, axis symmetric
side-walls. We have demonstrated that the consideration of the higher
order corrections to the stationary probability density leads to a
substantial improvement of the commonly employed Fick-Jacobs approach
towards extreme corrugate channels. 
The object of this work is to provide an analytic treatment to
biased Brownian motion in {\itshape cylindrical three-dimensional
  tubes} with periodically varying radius. 

In \secref{sec:structure} we introduce the model system: a Brownian
particle in a confined tube geometry with periodically modulated boundaries. The
central findings, namely the analytic expressions for the probability
density are presented in \secref{sec:longwave}. It turns out that the
latter scales linearly with the particle velocity derived within the
FJ approach, cf. \secref{sec:part_sinustube}. In
\secref{sec:corr} we calculate the particle mobility and the diffusion
coefficient and employ our analytical results to a tube with
sinusoidal varying cross-section. Section~\ref{sec:conclusion}
summarizes our findings. 

\section{Transport in confined structures}
\label{sec:structure}

\begin{figure}[t]
  \centering
  \includegraphics[width=\linewidth]{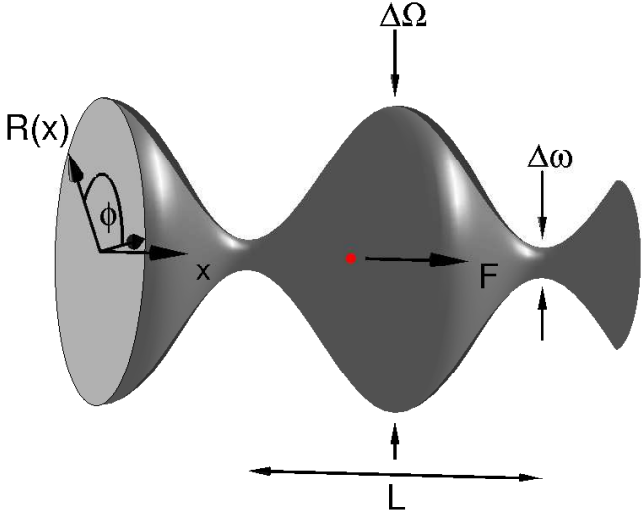}
  \caption{Sketch of a segment of a cylindrical tube with sinusoidally varying radius $R(x)$ that is
    confining the motion of the overdamped, point-like Brownian
    particle. The periodicity of the tube structures is $L$, the
    minimal and maximal tube widths are $\Delta \omega$ and $\Delta
    \Omega$, respectively. The constant force $F$ pointing in the
    direction of the tube is applied on the particles. }
  \label{fig:channelprob}
\end{figure}

The present paper deals with biased transport of overdamped Brownian
particles in a cylindrical tube with periodically varying
cross-section, respectively, radius $R(x)$. A sketch of a tube segment
with period $L$ is shown in \figref{fig:channelprob}. The particles
budge within a static fluid with constant friction coefficient
$\eta$.  As the particle radius (assumed to be point-like) is small
compared to the tube radius, hydrodynamic particle-particle
interactions as well as hydrodynamic particle-wall interactions can
safely be neglected. Further the particles are subjected to an
external force with static magnitude $F$ acting along the longitudinal
direction of the tube $\V{e}_x$, i.e. the corresponding potential is
$U(x,r,\phi)=-F\,x$.   

The evolution of the probability density $P\bracket{\V{q},t}$ of
finding the particle at the local position
$\V{q}=\bracket{x,r,\phi}^T$ at time $t$ is governed by the 
three-dimensional Smoluchowski equation
\cite{Risken,Haenggi1982}, i.e.,
\begin{subequations}
  \label{eq:sm}
  \begin{align}
    &\partial_t P\bracket{\V{q},t}+\nabla_{\V{q}} \cdot \V{J}\bracket{\V{q},t}=\,0\,, \label{eq:Smoluchowski}\\
    \intertext{where}
    &\V{J}\bracket{\V{q},t}=\,\frac{F}{\eta}\,P\bracket{\V{q},t}\,\V{e}_x-\frac{k_BT}
    {\eta}\,\nabla_{\V{q}}\,P\bracket{\V{q},t}\, \label{eq:def_probcurr}
  \end{align}
\end{subequations}                                                                                                                    
is the probability current
$\V{J}\bracket{\V{q},t}=\bracket{J^x,J^r,J^\phi}^T$ associated to the
probability density $P\bracket{\V{q},t}$. The Boltzmann constant is
$k_B$ and $T$ refers to the environmental temperature. At the tube
wall the probability current obeys the 
no-flux boundary condition (bc) caused by the impenetrability of the
tube walls, viz. $\V{J}\bracket{\V{q},t}\cdot\V{n}=0$ where $\V{n}$
is the out-pointing normal vector at the tube walls. For a tube with
radius $R(x)$ the bc read
\begin{subequations}
\begin{align}
 R'(x) J^x\bracket{\V{q},t}=&\,J^r\bracket{\V{q},t}\,,\quad \text{for} \quad r=R(x)\,. \label{eq:bcu}
\intertext{ The prime denotes the derivative with respect to x. As a
  result of symmetry arguments the probability current must be
  parallel with the tube's centerline at $r=0$} 
J^r\bracket{\V{q},t}|_{r=0}=&\,0\,. \label{eq:bcl}
\end{align} \label{eq:bc}
\end{subequations}
Further the probability density satisfies the normalization condition
$\int_{\mathrm{unit-cell}} P(\V{q},t)\, d^3\V{q} =1$
as well as the periodicity condition $P(x+m\,L,r,\phi,t)=P(x,r,\phi,t)\,,
\forall m \in \Z$. 

Since the external force acts only in longitudinal direction
$P\bracket{\V{q},t}$ is radial symmetric. This allows a reduction of
the problem's dimensionality from $3$D to $2$D by integrating
\eq{eq:Smoluchowski} over the angle $\phi$; yielding 
\begin{align}
 \partial_t
 P\bracket{x,r,t}=&\frac{k_{B}T}{\eta}\,\partial_x\left[e^{-\frac{U(x,r)}{k_BT}}\partial_x\bracket{
     e^{\frac{U(x,r)}{k_BT}}P(x,r,t)}\right] \notag \\ 
  &+\frac{k_{B}T}{\eta}\,\frac{1}{r}\partial_r\left[r\,\partial_r P(x,r,t)\right]\,,  \label{eq:Smoluchowski2}
\intertext{where the two-point probability density is defined as}
 P(x,r,t)=&\,\frac{1}{2\pi} \intl{0}{2\pi} d\phi\,P\bracket{x,r,\phi,t}\,. \label{eq:pdf_xr}
\end{align}

Integrating \eq{eq:Smoluchowski2} further over the cross-section and taking the boundary conditions Eqs.~\eqref{eq:bc}
into account, one gets
\begin{align}
 \partial_t P(x,t)=&\frac{k_{B}T}{\eta}\, \partial_x \intl{0}{R(x)} dr\,r \left[e^{-\frac{U(x,r)}{k_BT}}\partial_x\bracket{e^{\frac{U(x,r)}{k_BT}} P(x,r,t)}\right]. \label{eq:statSmo}
\intertext{Thereby the marginal probability density reads}
P(x,t)=&\frac{1}{2\pi} \intl{0}{R(x)}dr\,r\,\intl{0}{2\pi}d\phi\,P\bracket{x,r,\phi,t} \label{eq:marg_p} \,.
\end{align}

In Ref. \cite{Martens2011} we present a
perturbation series expansion in terms of a geometric parameter for
the problem of biased Brownian dynamics in a planar three-dimensional channel
geometry. Below we apply this method for Brownian motion in
cylindrical three-dimensional tubes. In doing so we introduce
dimensionless variables. We measure the longitudinal 
length in units of the period length $L$,
viz. $\overline{x}=x/L$. For the
rescaling of the 
$r$-coordinate, we introduce the dimensionless aspect parameter $\e$, i.e.
the difference of the widest cross-section 
of the tube, i.e. $\Delta\Omega$, and the most narrow constriction at the
bottleneck, i.e. $\Delta \omega$, in units of the period length,
yielding
\begin{align}
 \e=\frac{\bracket{\Delta\Omega-\Delta\omega}}{L}\,. \label{eq:def_eps}
\end{align}
The dimensionless parameter $\e$ characterizes the deviation of the boundary from the straight tube corresponding to $\e=0$.
Several authors considered different choices for the expansion parameter like the averaged half width \cite{Laachi2007,Yariv2007,Drazer2010} or
the ratio of an imposed anisotropy of the diffusion constants $\e^2=D_y/D_x$ \cite{Kalinay2006}. 

We next measure, for the case of finite corrugation
$\e \neq 0$, the radius $r$ in units of $\e L$, i.e. $r =\e L\,\overline{r}$ and, likewise, the boundary function $ R(x) = \e L\,h(x)$.
Time is measured in units of $\tau=\, L^2 \eta /(k_B\,T)$
which is twice the time the particle requires to overcome
diffusively, at zero bias $F=0$,  the distance $L$, i.e.
$\overline{t} = t / \tau$. The potential energy is rescaled by the
thermal energy $k_{\mathrm{B} } T$, i.e., for the considered
situation with a constant force component in longitudinal direction, one gets
$\overline{U}:=\overline{U}(x,r)= - F x / (k_{\mathrm{B}} T) = -f \overline{x}$, with
the dimensionless force magnitude \cite{Reguera2006, Burada2008}:
\begin{align}
 f=\frac{F\,L}{k_B\,T}\,.
\end{align}
After scaling the probability distribution reads
$p\bracket{\V{\overline{q}},\overline{t}}=\e^{2}\,L^3\,P\bracket{\V{q},t}$. Below we shall omit the overbar in our notation.

Further we concentrate only on the steady state, i.e., $\lim_{t \to  \infty} p(x,r,t):= p_\mathrm{st}(x,r)$, which is in
fact, the only state necessary for deriving the key quantities of particle transport like the average particle velocity $\av{\dot{x}}$
\begin{align}
  \av{\dot{x}}\equiv&\,\lim_{t\to\infty} \frac{\av{x(t)}}{t} =
  \intl{0}{1}dx\,\intl{0}{h(x)} dr\,r\,J_\mathrm{st}^x(x,r)\,,\label{eq:velocity}
\end{align}
 and the effective diffusion coefficient $D_\mathrm{eff}$ in force direction. The latter is given by
\begin{align}
 D_\mathrm{eff}=\lim_{t \to \infty} \frac{\av{x^2(t)}-\av{x(t)}^2}{2 t}\,, \label{eq:Deff}
\end{align}
and can be calculated by means of the stationary probability density
$p_{\mathrm{st}}\bracket{x,r}$ using an established method taken
from Ref.~\cite{Brenner}.

At steady state, the Smoluchowski equation \eq{eq:Smoluchowski2} in dimensionless units becomes
\begin{align}
  \e^2 \partial_x\left[e^{-U}\partial_x\bracket{ e^{U}p_\mathrm{st}(x,r)}\right]+\frac{1}{r}\partial_r\left[r\,\partial_r p_\mathrm{st}(x,r)\right]=\,0\,, \label{eq:Smoluchowski_scaled}
\end{align}
and the no-flux boundary conditions Eqs.~\eqref{eq:bc} read
\begin{subequations}
 \begin{align}
0=&\left[\partial_r p_\mathrm{st}(x,r)-\e^2 h'(x) e^{-U}\partial_x\bracket{e^{U} p_\mathrm{st}(x,r)}\right]_{r=h(x)}, \label{eq:bc2_u}\\
0=&\,\partial_r p_\mathrm{st}(x,r)|_{r=0}\,. \label{eq:bc2_l}
 \end{align} \label{eq:bc2}
\end{subequations}

\section{Asymptotic analysis}
\label{sec:longwave}

We apply the  asymptotic analysis
\cite{Yariv2007,Martens2011} to the problem stated by
\eq{eq:Smoluchowski_scaled} and Eqs.~\eqref{eq:bc2}. In doing so, we use for the
stationary probability density $p_{st}(x,r)$ (the index $st$ will be
omitted in the following) the ansatz
\begin{align}
 p(x,r)=\sum_{n=0}^{\infty} \varepsilon^{2n} p_n(x,r)\, \label{eq:prob_series}
\end{align}
in the form of a formal perturbation series in even orders of the
parameter $\varepsilon$. Substituting these expressions into
\eq{eq:Smoluchowski_scaled}, we find
\begin{align}
\begin{split}
0=\,\frac{1}{r}\partial_r\left[r \partial_r p_0(x,r)\right]+\sum_{n=1}^\infty \varepsilon^{2n}
\left\{\frac{1}{r}\partial_r\left[r \partial_r p_n(x,r)\right] \right.& \\ \left. +\partial_x\left[e^{-U} \partial_x\bracket{e^{U}p_{n-1}(x,r)}\right]\right\}\,.& \label{eq:longwave_pn}
\end{split}
\end{align}
The no-flux bc at the tube walls $r=h(x)$, cf. \eq{eq:bc2_u}, turns into
\begin{subequations}
\begin{align}
\begin{split}
0=\,\partial_r p_0(x,r)+&\sum_{n=1}^\infty \e^{2n}\left\{\partial_r p_{n}(x,r) \right. \\
 &\left. -h'(x)e^{-U} \partial_x\bracket{e^{U}p_{n-1}(x,r)}\right\}\,, \label{eq:longwave_bcu}
\end{split}
\intertext{and the bc at the centerline of the tube $r=0$, cf. \eq{eq:bc2_l}, then reads}
0=&\,\sum_{n=0}^\infty \e^{2n} \partial_r p_n(x,r)\,. \label{eq:longwave_bcl}
\end{align} \label{eq:longwave_bc}
\end{subequations}
We claim that the normalization condition for the probability density $p(x,r)$ corresponds to the zeroth solution $p_0(x,r)$ that is normalized to unity,
\begin{align}
 \av{p_0(x,r)}=\intl{0}{1}dx\,\intl{0}{h(x)}dr\,r\, p_0(x,r)=1\,. \label{eq:condnorm_pn}
\end{align}
Consequently the higher orders in the perturbation series have zero average, $\av{p_n(x,r)}=0\,,n=1,2,3,\ldots$ . Further each order $p_n$ has to obey the periodic boundary condition $p_n(x+m,r)=p_n(x,r)\,,\,\forall m \in \Z$.

In\secref{subsect:FJ}, we demonstrate that the zeroth order of the
perturbation series expansion coincides with the Fick-Jacobs
equation \cite{Jacobs,Zwanzig1992}. Referring to \cite{Stratonovich, Reguera2006} an expression for the average velocity $\av{\dot{x}}_0$ is known. Moreover, in \secref{subsect:higherorders}, the higher order corrections to the probability density are derived. Using those results
we are able to obtain corrections, see in \secref{subsect:mean_vel}, to
the average velocity beyond the zeroth order Fick-Jacobs approximation
presented in the next section.

\subsection{Zeroth Order:  the Fick-Jacobs equation}
\label{subsect:FJ}

For the zeroth order, \eq{eq:longwave_pn} read
\begin{subequations}
\begin{align}
 0=&\,\frac{1}{r}\partial_r\left[r \partial_r p_0(x,r)\right]
\intertext{supplemented with the corresponding no-flux boundary
condition at $r=0$ as well as at $r=h(x)$}
 0=&\,\partial_r p_0(x,r)\,.
\end{align}
\end{subequations}
We make the ansatz $p_0(x,r)=\,g(x)\,e^{-U}$ where $g(x)$ is an unknown function which has to be determined from
the second order $O\bracket{\varepsilon^2}$ balance given by
\eq{eq:longwave_pn}:
\begin{align}
0=&\,\partial_x\bracket{ e^{-U} g'(x)}+\frac{1}{r}\partial_r \left[r\,\partial_r\bracket{p_1(x,r)}\right].
\end{align}
Integrating the latter over the radius $r$ and taking the no-flux boundary conditions Eqs.~\eqref{eq:longwave_bc}
into account, one immediately obtains
\begin{align}
 0=&\,\partial_x \bracket{e^{-A(x)} g'(x)}\,. \label{eq:def_g}
\end{align}
The effective entropic potential $A(x)$ is defined by
\begin{align}
 e^{-A(x)}=&\,\intl{0}{h(x)} dr\,r\,e^{-U(x,r)}\,, \label{eq:def_effpot}
\intertext{and for the problem at hand the latter looks explicitly}
A(x)=&\,-f\,x-\ln\bracket{\frac{h^2(x)}{2}}\,.
\end{align}
Note, that upon an irrelevant additive constant, i.e. $\ln(2\pi)$, the effective entropic
potential corresponds to that given in Ref.~\cite{Jacobs}.

The normalized stationary probability density within the zeroth order reads \cite{Risken}
\begin{align}
 p_0(x,r)=e^{-U}g(x)=&\frac{e^{-U(x,r)}\intl{x}{x+1}
   e^{A(x')}dx'}{\intl{0}{1}dx e^{-A(x)}\intl{x}{x+1}
   e^{A(x')}dx'}\,, \label{eq:finalsol_p0}
\intertext{and, moreover, the marginal probability density \eq{eq:marg_p} becomes}
p_0(x)=&\,e^{-A(x)}\,g(x)\,.  \label{eq:finalsol_projp0}
\end{align}
Expressing next $g(x)$ by $p_0(x)$, see \eq{eq:def_g}, then yields
the celebrated stationary Fick-Jacobs equation
\begin{align}
 0=\partial_x\left[e^{-A(x)}\partial_x \bracket{e^{A(x)}\,p_0(x)}\right] \label{eq:FJ_pdf}
\end{align}
derived  previously in Ref.~\cite{Zwanzig1992,Reguera2001}. 

Thus, we demonstrate that the leading order term of the asymptotic analysis is equivalent to the FJ-equation. In the FJ equation the problem of biased Brownian dynamics in a confined $3$D geometry is replaced by Brownian motion in the tilted periodic one-dimensional potential $A(x)$. In general, the stationary probability density of finding an overdamped Brownian particle budging in a tube with periodically varying cross-section is sufficiently
described by \eq{eq:FJ_pdf} as long as the extension of the bulges of the
tube structures is small compared to the periodicity, i.e. $\e \ll 1$.

Then the average particle current is calculated by integrating the probability flux
$J_0^x$ over the unit-cell \cite{Stratonovich,Ivanchen,Tikhonov}
\begin{align}
 \av{\dot{x}(f)}_0=&\,\intl{0}{1}dx\,\intl{0}{h(x)}dr\,r\,J_0^x(x,r)
  \nonumber \\
=&\,\frac{1-e^{-f}}{\intl{0}{1}dx\,e^{A(x)}\,\intl{x-1}{x}
  e^{-A(x')}\,dx'}\,. \label{eq:currFJ}
\end{align}
In the spirit of linear response theory, the mobility in dimensionless
units is defined by the ratio of the mean
particle current \eq{eq:currFJ} and the applied force $f$, yielding
\begin{align}
 \mu_0\bracket{f}=\frac{\av{\dot{x}(f)}_0}{f}\,. \label{eq:def_mob}
\end{align}
Note, that in order to obtain the mobility in physical units one has
to multiply $\mu_{0}$ with the mobility of unconfined particles,
i.e. $1/\eta$.

Further, resulting from the normalization condition \eq{eq:condnorm_pn}, the average particle velocity \eq{eq:velocity} simplifies to
\begin{align}
 \av{\dot{x}}=&\av{\dot{x}}_0-\sum_{n=1}^\infty\e^{2n}\av{\partial_x p_n(x,r)}\,. \label{eq:series_vx}
\end{align}
Therefore we derive that the average particle current is composed of (i) the
Fick-Jacobs result $\av{\dot{x}}_0$ , cf. \eq{eq:currFJ}, and (ii) becomes
corrected by the sum of the averaged derivatives of the higher
orders $p_n(x,r)$. We next address the higher order corrections $p_n(x,r)$ of the probability
density which become necessary for more corrugated structures.

\subsection{Higher order contributions to the Fick-Jacobs equation}
\label{subsect:higherorders}

According to \eq{eq:longwave_pn}, one needs to iteratively solve
\begin{align}
\label{eq:iterative}
 \frac{1}{r}\partial_r\left[r\partial_r p_n(x,r)\right]=&\,\mathfrak{L}\,p_{n-1}(x,r)\,,\quad n\geq 1\,,
\end{align}
under consideration of the boundary conditions Eqs.~\eqref{eq:longwave_bc}.
In \eq{eq:iterative}, we make use of the operator $\mathfrak{L}$,
reading $\mathfrak{L}=\bracket{f\,\partial_x-\partial_x^2}$. 
Each solution of the second order partial differential equation
\eq{eq:iterative} possesses two integration constants $d_{n,1}$ and
$d_{n,2}$. The first one, $d_{n,1}$, is determined by the no-flux
bc at the centerline $r=0$, cf. \eq{eq:longwave_bcl}, while the second provides
the zero average condition $\av{p_n(x,r)}=0\,,\,n\geq 1$.

For the first order correction, the determining equation reads
\begin{align}
 \frac{1}{r}\partial_r\left[r\partial_r p_1(x,r)\right]=&\,2\,\av{\dot{x}}_0\,\partial_x
 \bracket{\frac{1}{h^2(x)}}\,,
\intertext{and after integrating twice over $r$, we obtain}
 p_1(x,r)=&\,-\av{\dot{x}}_0\,\bracket{\frac{h'(x)}{h^3(x)}}\,r^2\,. \label{eq:finalsol_p1}
\end{align}
Hereby, as requested above, the first integration
constant $d_{1,1}(x)$ is set to $0$ in order to fulfill the no-flux
bcs, and the second must provide the normalization
condition \eq{eq:condnorm_pn}, i.e. $d_{1,2}=0$. One notices that the
first correction to the probability density becomes positive if the
confinement is constricting, i.e. for $h'(x) <0$ and
$\av{\dot{x}}_0\neq 0$. In contrast, the probability density becomes
less in unbolting regions of the confinement, i.e. for $h'(x)>0$.
Please note, that the first order correction scales linearly with
the average particle current $\av{\dot{x}}_0$. Overall, the break of
spatial symmetry observed within numerical simulations in previous
works \cite{Burada2007,Burada2010_PRE} is reproduced by this very
first order correction. Particularly, with increasing forcing, the
probability for finding a particle close to the constricting part of
the confinement increases, cf. Ref.~\cite{Burada2007,Burada2010_PRE}.

Upon recursively solving, the higher order corrections, $n \geq 1$, are given by
\begin{align}
 p_n\bracket{x,r}=&\,\mathfrak{L}^{n} p_{0}(x,r)\,\frac{r^{2n}}{\bracket{2^{n}n!}^2}+d_{n,2}\,, \label{eq:finalsol_pn} \\ 
=&\,2\,\av{\dot{x}}_0\,\frac{r^{2n}}{\bracket{2^{n}n!}^2}\,\mathfrak{L}^{n-1} \partial_x\bracket{\frac{1}{h^2(x)}}+d_{n,2} \notag
\intertext{where the operator $\mathfrak{L}$ applied $n$-times yields the expression}
\mathfrak{L}^{n}=&\,\sum_{k=0}^{n} \begin{pmatrix} n \\ k \end{pmatrix}
\bracket{-1}^k f^{n-k}\,\p{ ^{n+k}}{x^{n+k}}. \label{eq:Ln}
\end{align}
Note that each single order $p_n(x,r)$, cf. \eq{eq:finalsol_pn}, satisfies the normalization condition, the bc at the centerline but does not obey the bc at the tube wall, see \eq{eq:longwave_bcu}. The stationary probability density $p(x,r)$ is obtained by summing all correction terms, cf. \eq{eq:prob_series}, yielding
\begin{align}
 p(x,r)=p_0(x,r)+\sum_{n=1}^\infty \e^{2n}\bracket{\mathfrak{L}^n
   p_0(x,r) \frac{r^{2n}}{\bracket{2^{n}n!}^2}+d_{n,2}}. \label{eq:full_probdens}
\end{align}
Inserting \eq{eq:full_probdens} into the equation
for the no-flux bc at the tube wall, cf. \eq{eq:bc2_u}, results to 
\begin{align}
0\equiv&\sum_{n=0}^\infty \e^{2n+2} \partial_x \intl{0}{h(x)}dr\,r\,e^{-U(x,r)}\partial_x\bracket{e^{U(x,r)} p_n(x,r)}\notag \\
=&\e^2 \partial_x \intl{0}{h(x)}dr\,r\,e^{-U(x,r)}\partial_x\bracket{e^{U(x,r} p(x,r)}\,.
\end{align}
According to \eq{eq:statSmo} the latter equals zero in the steady
state. 

Summing up, the exact solution for the stationary probability density of finding a biased Brownian particle in tube is given by \eq{eq:full_probdens}. The latter solves the corresponding Smoluchowski equation \eq{eq:Smoluchowski_scaled} under satisfaction of the normalization as well as the periodicity requirements. More importantly the solution, cf. \eq{eq:full_probdens}, obeys the no-flux boundary conditions at the centerline $r=0$ as well as at the tube wall $r=h(x)$. Further one notices that $p(x,r)$ is fully determined by the Fick-Jacobs results $p_0(x,r)$. Caused by $\mathfrak{L}\,p_0(x,r)\propto \av{\dot{x}(f)}_0$,
the contribution of the higher order corrections to the $2$D
probability density scales linear with the average particle current in
the FJ limit $\av{\dot{x}}_0$, cf. \eq{eq:finalsol_pn}. The latter is
determined by the break of spatial symmetry induced by the external
force. Consequently, in the absence of the external force $f=0$ the
stationary probability density equals the zeroth order contribution
$p(x,r)=p_0(x,r)=const$ despite the value of $\e$. 

According to \eq{eq:series_vx}, one recognizes that the average particle
current scales with the average particle current obtained from the Fick-Jacobs formalism $\av{\dot{x}}_0$ for
all values of $\e$. Therefore, in order to validate the obtained results for $p(x,r)$ as well as to derive  correction to the mean particle current it is required to calculate $\av{\dot{x}}_0$ first.

\section{Transport quantities for a sinusoidally shaped tube}
\label{sec:part_sinustube}

In the following we study the key transport quantities like the particle mobility $\mu(f)$ and the effective diffusion coefficient $D_\mathrm{eff}(f)$ of point-like Brownian particles moving in a sinusoidally-shaped\cite{Burada2008} tube. The dimensionless boundary function
$h(x)$ reads
\begin{align}
 h\bracket{x}=&\,\frac{1}{4}\bracket{\frac{1+\delta}{1-\delta}+\sin\bracket{2\pi\,x}}\,, \label{eq:conf}
\end{align}
and is illustrated in \figref{fig:channelprob}. The function $h(x)$ is solely governed by the aspect ratio of the minimal and maximum tube width $\delta=\Delta\omega/\Delta\Omega$. Obviously
different realizations of tube geometries can possess the same
value of $\delta$. The number of  orders have to taken into account in
the perturbation series \eq{eq:prob_series}, respectively, the
applicability of the {\it Fick-Jacobs approach} to the problem, depends
only on the value of the geometric parameter $\e=\Delta\Omega\bracket{1-\delta}/L$ for a given aspect ratio
$\delta$.

First, we obtain the particle mobility $\mu_0$ within the zeroth-order
(Fick-Jacobs approximation). Referring to Eqs.~\eqref{eq:currFJ} and
\eqref{eq:def_mob} the dimensionless mobility is given by
\begin{align}
 \mu_0(f)=&\,\frac{1-e^{-f}}{f\,\intl{0}{1}dx\, e^{-fx'}/h^2(x')\,\intl{x-1}{x}e^{fx}h^2(x)\,dx'}\,.
\end{align}
For the considered sinusoidal boundary function, cf.~\eq{eq:conf}, we obtain
\begin{align}
\begin{split}
 \frac{1}{\mu_0(f)}=&\,\frac{1}{2\sqrt{b^2-1}^3}\left\{ b\bracket{2b^2+1}-\frac{4b\,f^2}{f^2+\bracket{2\pi}^2} \right. \\  &\left.+\frac{\left(2\sqrt{b^2-1}^3-2b^3+3b\right)\,f^2}{f^2+\bracket{4\pi}^2}\right\},
\end{split}\label{eq:mob_FJ}
\end{align}
with the substitute $b=(1+\delta)/(1-\delta)$. Caused by the reflection symmetry of the boundary function $h(x)$ the particle mobility obeys $\mu_0(-f)=\mu_0(f)$. Thus it is sufficient to discuss only the behavior for $f\geq 0$. 

In the limiting case of infinite large force strength, the mobility goes to
\begin{align}
\lim_{f\to \infty} \mu_0(f)=&1\,.
\intertext{With decreasing force magnitude $f$ the mobility decreases as well till $\mu_0(f)$ attains the asymptotic value}
 \lim_{f\to 0} \mu_0(f)=&\,\frac{2 \sqrt{\delta}}{1+\delta}\,\frac{8\,\delta}{3\delta^2+2\delta+3}\,.\label{eq:mobdeff_FJ_limf0}
\end{align}
In the diffusion dominated regime, $|f|\ll 1$, the Sutherland-Einstein relation
emerges~\cite{Burada2009_PTRSA,chaos05} and thus the
dimensionless mobility equals the dimensionless effective diffusion coefficient:
\begin{align}
  \label{eq:SEinstein}
  \lim_{f\to0}\mu(f) = \lim_{f\to 0} D_{\mathrm{eff}}(f).
\end{align}
In the limit of vanishing bottleneck width, i.e. $\delta \to 0$, the mobility, respectively, $D_\mathrm{eff}$ tends to $0$. In contrast, for straight tubes
corresponding to $\delta=1$, i.e. $\e = 0$, the mobility as well as the effective diffusion coefficient equal their free values which are one in the considered scaling. 

In a recent work \cite{Martens2011}, we studied the biased Brownian
motion in a planar three-dimensional channel geometry with 
periodically varying width $2h(x)$. We found that the
asymptotic value for the mobility is given by
$2\sqrt{\delta}/\bracket{1+\delta}$ for $f\to0$, cf. Eq.~(45) in
Ref.~\cite{Martens2011}. Comparing this result with the asymptotic
value \eq{eq:mobdeff_FJ_limf0} one notices that the mobility
and the effective diffusion coefficient in a tube, respectively, are less
compared to case of a planar channel geometry.

\begin{figure}[t]
\includegraphics[width=\linewidth]{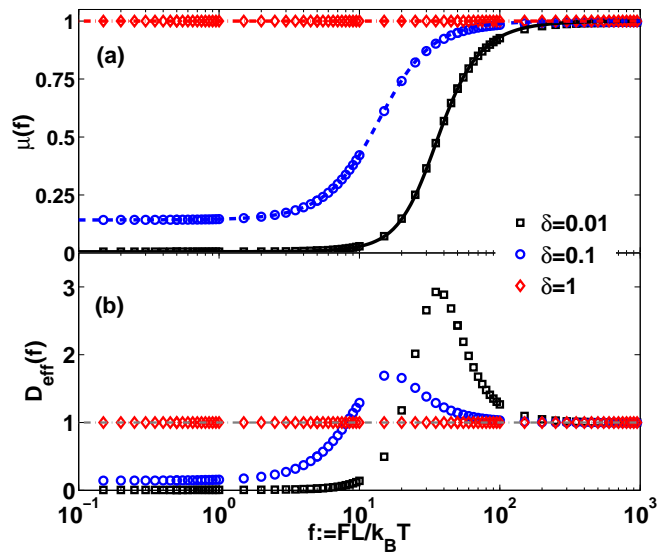}
\caption{(Color online) The particle mobility (a) and the effective diffusion
  constant (b) for a Brownian particle moving inside a sinusoidal tube
  are depicted as function of the external force magnitude $f$. The maximum tube width is kept fixed, viz. $\Delta\Omega=0.1$, while the aspect ratio is varied $\delta = 0.01,0.1,1$, respectively, the corresponding values for $\e$ are $\e=0.099,0.09,0$. The symbols correspond to the numerical obtained mobility, respectively, the
  effective diffusion coefficient. In subfigure (a) the lines correspond to the analytic result, cf.~\eq{eq:mob_FJ}.}
\label{fig:mobdeff}
\end{figure}

In \figref{fig:mobdeff} we depict the dependence of the mobility $\mu(f)$ and the effective diffusion coefficient $D_\mathrm{eff}(f)$ on the external force magnitude $f$. The numerical results are obtained by solving the stationary Smoluchowski equation \eq{eq:Smoluchowski}
using finite element method \cite{FreeFem} and subsequently calculating the average particle current according to \eq{eq:velocity}. In order to
determine the effective diffusion coefficient $D_\mathrm{eff}(f)$, one has to solve numerically the reaction-diffusion equation for the {\itshape
B-field} \cite{Brenner,Laachi2007}. 

Referring to \figref{fig:mobdeff}(a) one notices that the analytic predictions for the particle mobility \eq{eq:mob_FJ} are corroborated by numerics. Further, one observes that for the case of smoothly varying tube geometry, i.e. $\Delta\Omega/L
\ll 1$, the analytic result is in excellent agreement with the
numerics for a large range of dimensionless force magnitudes $f$,
indicating the applicability of the Fick-Jacobs approach. As long as
the extension of the bulges of the tube structures is small compared to the periodicity, sufficiently fast transversal equilibration, which serves as fundamental ingredient for the validity of the Fick-Jacobs approximation is taking place.

The effective diffusion coefficient $D_\mathrm{eff}(f)$ exhibits a non-monotonic dependence
versus the dimensionless force $f$, see \figref{fig:mobdeff}(b). It starts out with a value that is less than the free diffusion constant in the diffusion dominated regime,i.e. $|f|\ll 1$. According to the Sutherland-Einstein-relation the value equals the mobility value, cf.~\eq{eq:mobdeff_FJ_limf0}.  Then it reaches a maximum with increasing $f$ and finally approaches the value of the free diffusion from above. Further one notices that the location of the diffusion peak as well as the peak height depends on the aspect ratio $\delta$. With decreasing width at the bottleneck, while keeping the maximum width $\Delta\Omega$ fix, the diffusion peak is shifted towards larger force magnitude $f$. Simultaneously the peak height grows. In the limit of a straight tube, i.e. $\delta\to 1$, as expected the effective diffusion coefficient coincide with its free value which is one in the considered scaling.

\begin{figure}
\includegraphics[width=\linewidth]{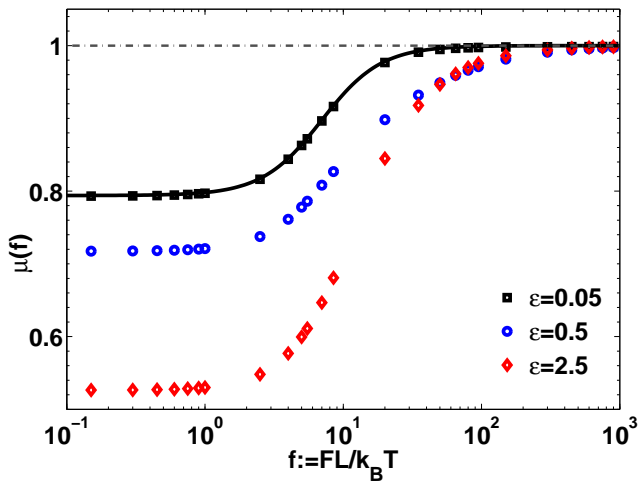}
\caption{(Color online) The influence of the geometric parameter $\e$ on the particle mobility is presented. The value of $\e$ is varied, viz. $\e=0.05,0.5,2.5$ , while the aspect ratio is kept fixed, viz. $\delta=0.5$. The solid line corresponds to the analytic result \eq{eq:mob_FJ} while the dash-dotted line indicates the asymptotic value one.}
\label{fig:mobdeff_eps}
\end{figure}

In \figref{fig:mobdeff_eps} we present the impact of the expansion parameter $\e$ on the particle mobility. It turns out that for values of $\e \lesssim 0.1$ the Fick-Jacobs approximation is in very good agreement with the simulation. With increasing geometric parameter $\e$ the difference between the FJ-result (solid line) and the numerics is growing. The more available space in the tube leads to a decrease of the particle mobility, respectively, of the effective diffusion coefficient. Consequently, the higher order corrections to the stationary probability density $p(x,r)$, see \eq{eq:prob_series}, respectively, the corrections to the mobility, cf.~\eq{eq:series_vx}, need to be included in order to provide a better agreement.

\section{Corrections to the mobility and diffusion coefficient}
\label{sec:corr}

A common used way to include the corrugation of the channel structure
bases on the concept of the spatially dependent diffusion coefficient
$D(x,f)$ which was introduced by Zwanzig \cite{Zwanzig1992} and
subsequently supported by the study of Reguera and Rubi
\cite{Reguera2001}. Zwanzig obtained the FJ equation, cf. \eq{eq:FJ_pdf}, from the full $3$D Smoluchowski equation upon eliminating the transverse degrees of
freedom supposing infinitely fast relaxation. In a more detailed view, we have to notice that diffusing
particles can flow out from/ or towards the wall in $r$-direction only at finite time. These finite relaxation processes are included by scaling the diffusion constant in longitudinal direction by a position dependent function, viz. $D(x,f)$. The latter substitutes the constant diffusion coefficient - which is one in the considered scaling - in the common stationary FJ equation, cf. \eq{eq:FJ_pdf}, yielding
\begin{align}
 0=\,\partial_x\left[D\bracket{x,f} e^{-A\bracket{x}}\partial_x\bracket{e^{A\bracket{x}}
 p\bracket{x}}\right]. \label{eq:FJ_pdf_spatialD}
\end{align}
According to \eq{eq:FJ_pdf_spatialD} an expression for the particle mobility that is similar to the Stratonovich formula \cite{Stratonovich, Reguera2006} for the
mobility in titled periodic energy landscapes, but now including the the spatial diffusion coefficient $D(x,f)$, can be derived \cite{Reguera2006,Burada2007}. In the diffusion dominated regime, i.e. $|f|\ll 1$, this expression simplifies to the Lifson-Jackson formula \cite{Lifson1962,Burada2007}
\begin{align}
 \lim_{f\to 0} \mu(f)=\lim_{f\to 0}  D_\mathrm{eff}(f)=\frac{1}{\av{h^2(x)}\av{\frac{1}{D(x,0)\,h^2(x)}}}\,, \label{eq:mobdeff_f0_Dx}
\end{align}
with the period average $\av{\cdot}=\int_{0}^{1} \cdot dx$. Unfortunately, for many boundary functions $h(x)$ it is impossible to analytically evaluate the expression \eq{eq:mobdeff_f0_Dx}.

\subsection{Spatially dependent diffusion coefficient $D(x,f)$}
\label{subsect:spatD}

A first systematically treatment taking the finite diffusion time into account was presented by
Kalinay and Percus (KP) \cite{Kalinay2005,Kalinay2006}. Their suggested mapping procedure enables the
derivation of higher order corrections in terms of an
expansion parameter $\e_{KP}^2$, which is the ratio of the diffusion
constants in the longitudinal and transverse directions. Within this scaling, KP have shown that the fast transverse modes (transients) separate from the slow longitudinal ones and therefore the transients can be projected out by integration over the transverse directions.

In what follows we present a derivation for the spatially dependent
diffusion coefficient $D(x,f)$ which based on our previously considered perturbation series expansion for the stationary probability density \secref{sec:longwave}. According to \eq{eq:statSmo} the marginal probability
current $J^{x}(x)$, equivalent to \eq{eq:FJ_pdf_spatialD}, can be derived in an alternative way using the stationary two-point probability density $p(x,r)$, yielding
\begin{align}
\begin{split}
 -J^x(x)=&\,D\bracket{x,f} e^{-A\bracket{x}} \partial_x\bracket{e^{A\bracket{x}}
 p\bracket{x}} \\ =&\,\intl{0}{h(x)}dr\,r\, e^{-U\bracket{x,r}} \partial_x\bracket{e^{U\bracket{x,r}}
 p\bracket{x,r}}\,. \label{eq:dspatial_1}
\end{split}
\end{align}
The second equality determines the sought-after spatial dependent
diffusion coefficient $D(x,f)$. 

One immediately notices that the relation \eq{eq:dspatial_1} simplifies to $D(x,f)\,f\,p(x)=\int_{0}^{h(x)}dr\,r\,f\,p(x,r)$ in the force dominated regime $|f|\gg 1$. Then it follows that the spatially diffusion coefficient equals the free one, which is one in the considered scaling,
\begin{align}
 \lim_{f\to\infty} D(x,f)=1\,.
\end{align}

In the opposite limit of small force strengths, i.e. for $|f| \ll 1$, diffusion is the dominating process. Then \eq{eq:dspatial_1} simplifies to
\begin{align}
 D\bracket{x,f} h^2(x)\partial_x\bracket{\frac{p\bracket{x}}{h^2(x)}}=\intl{0}{h(x)}dr\,r\, \partial_x p\bracket{x,r}\,. \label{eq:dspatial_2}
\end{align}
Inserting our result for the stationary probability density $p(x,r)$, cf. \eq{eq:full_probdens}, into \eq{eq:dspatial_2} we determine an expression for $D(x,f)$. In compliance with Ref.~\cite{Kalinay2006}, we
make the ansatz that all but the first derivative of the boundary
function $h(x)$ are negligible. Then the
$n$-times applied operator $\mathfrak{L}$, cf. \eq{eq:Ln},
simplifies to $ \mathfrak{L}^{n}=\,(-1)^n\,\p{ ^{2\,n}}{x^{2\,n}}$, yielding,
\begin{align}
p_n(x,r)=\,2\,\bracket{-1}^n\av{\dot{x}}_0 \frac{\bracket{2n}!}{\bracket{2^{n}n!}^2}\frac{(h')^{2n-1}}{h^{2n+1}}r^{2n}+O(h''(x))\,.
\end{align}
Inserting the latter into \eq{eq:dspatial_1} and calculating the complete sum, one finds
\begin{align}
 \lim_{f\to 0} D(x,f)= &\,\frac{1}{\sqrt{1+\bracket{\e h'(x)}^2}}+O\bracket{h''(x)} \label{eq:deff_kalinay_series}
\end{align}
for the spatially dependent diffusion coefficient $D(x,f)$ in the
diffusion dominated regime, i.e. for $|f|\ll 1$. Using our above presented series expansion for the stationary probability density $p(x,r)$ we confirm the expression for the spatially dependent diffusion coefficient $D(x,f)$ previously derived by Reguera and Rubi \cite{Reguera2001} and KP \cite{Kalinay2005,Kalinay2006}.

\subsection{Corrections based on perturbation series expansion}
\label{subsect:mean_vel}

Next, we derive an estimate for the mean particle current $\av{\dot{x}(f)}$
based on the higher expansion orders $p_n(x,r)$. Referring to \eq{eq:series_vx}, the average particle current is composed of (i) the Fick-Jacobs result $\av{\dot{x}}_0$ , cf. \eq{eq:currFJ}, and (ii) becomes
corrected by the sum of the averaged derivatives of the higher
orders $p_n(x,r)$. Immediately one notices that the integration constant $d_{n,2}$, resulting from the normalization condition \eq{eq:condnorm_pn}, does not influence the result for the average particle velocity, cf.~\eq{eq:series_vx}.

We concentrate on the diffusion dominated limit $|f|\ll 1$ and further we make the ansatz that all but the first derivative of the boundary function $h(x)$ are negligible \cite{Kalinay2006,Martens2011}. Then the partial derivative of $p_n(x,r)$ with respect to $x$ simplifies to
\begin{align}
\partial_x p_n(x,r)=\,2\av{\dot{x}}_0(-1)^{n+1}
\frac{(h')^{2n}}{h^{2n+2}} \frac{\bracket{2n+1}!\,r^{2n}}{\bracket{2^{n}n!}^2}+O(h''(x))\,.
\end{align}
Inserting the latter into \eq{eq:series_vx} and integrating the results over one unit-cell, results in
\begin{align}
 &\lim_{f\to 0}\av{\dot{x}(f)}\simeq\lim_{f\to 0}\,\av{\dot{x}(f)}_0\,\sum_{n=0}^\infty \frac{\bracket{-1}^n \bracket{2n+1}!}{2^{2n+1}\bracket{n!}\bracket{n+1}!} \av{\bracket{\e h'(x)}^{2n}} \notag \\
&\simeq\lim_{f\to 0}\,\av{\dot{x}(f)}_0\av{\frac{2}{\bracket{\e h'(x)}^2}\bracket{1-\frac{1}{\sqrt{1+\bracket{\e h'(x)}^2}}}}. \label{eq:mobdeff_gen_limf0}
\end{align}
We obtain that the average transport velocity is obtained as the product of the zeroth-order Fick-Jacobs result and the expectation
value of a complicated function including the slope of the boundary for $|f|\ll 1$. In the previously studied case of biased Brownian motion in a $3$D planar channel geometry \cite{Martens2011} we have found that the average transport velocity is obtained as the product of the zeroth-order Fick-Jacobs result and the expectation value of the spatially dependent diffusion coefficient $\av{D(x,0)}$. In contrast to the $3$D planar geometry,  for biased transport in extreme corrugated tubes the corrections term to the particle velocity does not coincide with the expectation value of $D(x,0)$, see \eq{eq:deff_kalinay_series}. 

Calculating the expectation value in \eq{eq:mobdeff_gen_limf0} for the
considered tube geometry, cf. \eq{eq:conf}, yields to the estimate
\begin{align}
 \lim_{f\to 0}\av{\dot{x}(f)}\simeq&\lim_{f\to 0}\,\av{\dot{x}(f)}_0\,_2F_1\bracket{\frac{1}{2},\frac{1}{2},2,-\bracket{\frac{\e\pi}{2}}^2},
\label{eq:mob_limf0}
\end{align}
where $_2F_1\bracket{\cdot}$ is the first hypergeometric function. We derive that in the diffusion dominated regime the average velocity is obtained as the
product of the zeroth-order Fick-Jacobs result and the correction term $_2F_1\bracket{\cdot}$ including the corrugation of the tube structure. Referring to the Sutherland-Einstein relation, cf. \eq{eq:SEinstein}, if the average current $\av{\dot{x}(f)}_0$ (or the
effective diffusion coefficient $D_\mathrm{eff}^0(f)$) is known in the
zeroth order, the higher order corrections to both quantities can be
obtained according to \eq{eq:mob_limf0}.  

We have to emphasize that considering only the first derivative of the boundary function $h'(x)$ results in an additional term proportional to $\e^2 \,_2F_1\bracket{3/2,3/2,3,-\bracket{\e\pi/2}^2}$. Taking further the second derivative $h''(x)$ into accounts indicates that this second term is negligible compared to $_2F_1\bracket{1/2,\ldots}$ for arbitrarily value of $\e$.

\begin{figure}[t]
\includegraphics[width=\linewidth]{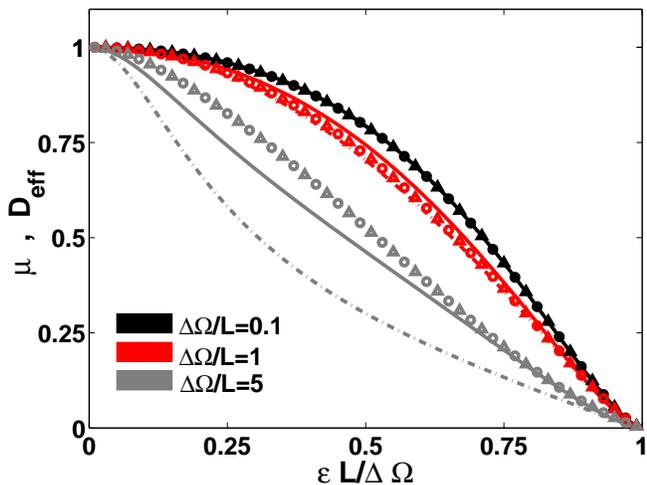}
\caption{(Color online) Comparison of the analytic theory versus precise
  numerics (in dimensionless units): The mobility and the effective diffusion
  constant for a Brownian particle moving inside a tube with sinusoidal varying radius
  are depicted as function of geometric parameter $\e$ in units of the maximum channel width $\Delta\Omega$. The latter is varied $\Delta\Omega/L = 0.1,1,5$ (from top to bottom) while the external bias is kept fixed $f=10^{-3}$ (corresponding to the diffusion dominated
  regime). The symbols correspond to the numerical obtained mobility (triangles) and the
  effective diffusion coefficient (circles). The solid lines correspond to
  analytic higher order result, cf.~\eq{eq:mob_limf0}. The zeroth  order - Fick-Jacobs results given by \eq{eq:mobdeff_FJ_limf0}
  collapse to a single curve hidden by the solid line for $\Delta\Omega/L=0.1$. In addition the numerical evaluation of \eq{eq:mobdeff_f0_Dx} is represented by the dash-dotted lines.}
\label{fig:mobdeffperi_eps}
\end{figure}

In \figref{fig:mobdeffperi_eps}, we present the dependence of the $\mu(f)$ (triangles) and
$D_\mathrm{eff}(f)$ (circles) on the slope parameter $\e$
for $f=10^{-3}$. One observes that the numerical results for the effective diffusion coefficient $D_{\mathrm{eff}}(f)$ and the mobility $\mu(f)$ coincide for all values of $\e$, thus corroborating the Sutherland-Einstein relation. In addition, the Fick-Jacobs result, given by
\eq{eq:mobdeff_FJ_limf0}, the higher order result (solid lines), see \eq{eq:mob_limf0}, and the numerical evaluation of the Lifson-Jackson formula using $D(x,0)$ (dash-dotted lines), cf. \eq{eq:mobdeff_f0_Dx}, are depicted in \figref{fig:mobdeffperi_eps}.

For the case of smoothly varying tube geometry, i.e. $\Delta\Omega/L
\ll 1$, all analytic expressions are in excellent agreement
with the numerics, indicating the applicability of the Fick-Jacobs approach. In virtue of \eq{eq:def_eps}, the geometric parameter is defined by $\e=\bracket{\Delta\Omega-\Delta \omega}/L$ and hence the maximal value of $\e$ equals $\Delta\Omega/L$. Consequently the influence of the higher expansion orders $\e^{2n}\av{\partial_x p_n(x,r)}$ on the average velocity \eq{eq:series_vx} and on the mobility, respectively, becomes negligible if the maximum tube's width $\Delta\Omega/L$ is small.

With increasing maximum width the difference between the
FJ-result \eq{eq:mobdeff_FJ_limf0} and the numerics is growing. Specifically, the FJ-approximation overestimates the mobility $\mu$
and the effective diffusion coefficient $D_{\mathrm{eff}}$. Consequently the corrugation of the tube geometry
needs to be included. The consideration of $D(x,0)$, cf. \eq{eq:mobdeff_f0_Dx} provides a good agreement for a wide range of $\e$-values as long as the maximum width $\Delta\Omega/L$ is on the scale to the period length of
the tube, i.e. $\Delta\Omega/L\sim 1$. Upon further increasing the maximum width $\Delta\Omega/L$ diminishes
the range of applicability of the presented concept. In detail the expression \eq{eq:mobdeff_f0_Dx} drastically underestimate the numerical results due to the neglect of the higher derivatives of the boundary function $h(x)$. Put differently, the higher derivatives of $h(x)$ become significant for $\Delta\Omega/L \gtrsim 1$.

In contrast, one notices that the result for the mobility and the diffusion coefficient, respectively, based on the higher order corrections to the stationary probability density \eq{eq:mob_limf0} is in very good agreement with the numerics. For tube geometries where the maximum width $\Delta\Omega/L$ is on the scale to the period length, i.e. $\Delta\Omega/L\sim 1$, the correction estimate matches perfectly with the numerical results. Further increase of the tube width results in a small deviation from the simulation results.

\section{Summary and conclusion}
\label{sec:conclusion}

In summary, we have considered the transport of point-sized Brownian particles under the
action of a constant and uniform force field through a $3$D tube. The cross-section, respectively, the radius of the tube varies periodically.

We have presented a systematic treatment of particle transport by
using a perturbation series expansion of the stationary probability density in terms of a smallness parameter which
specifies the corrugation of the tube walls. In particular, it
turns out that the leading order term of the series expansion is
equivalent to the well-established {\itshape Fick-Jacobs approach}
\cite{Jacobs,Zwanzig1992}. The higher order corrections to the
probability density become significant for extreme bending of the
tube's side-walls. Analytic results for each order of the perturbation
series have been derived. Similar to biased Brownian motion in a $3$D planar channel all higher
order corrections to the stationary probability density scale with the average
particle current obtained from the Fick-Jacobs formalism.

Moreover, for the diffusion dominated regime, i.e. for small forcing $|f|\ll 1$, we calculate the correction to the mean particle velocity originated by the tube's corrugation using the series expansion for the stationary probability density. According to the
Sutherland-Einstein relation, the obtained relation is also valid for
the effective diffusion coefficient. In addition, by using the higher order
corrections we present an alternative derivation for
the spatially dependent diffusion coefficient $D(x,f)$ which
substitutes the constant diffusion coefficient present in the common
Fick-Jacobs equation based on similar assumptions as those suggested
by Kalinay and Percus as well as by Rubi and Reguera.

Finally, we have applied our analytic results to a specific
example, namely, the particle transport through a tube with
sinusoidally varying radius $R(x)$. We corroborate our theoretical
predictions for the mobility and the effective diffusion coefficient
with precise numerical results of a finite element calculation of the
stationary Smoluchowski-equation. 

In conclusion, the consideration of the higher order corrections leads to a substantial improvement of the
Fick-Jacobs-approach, which corresponds to the zeroth order in our
perturbation analysis, towards more winding boundaries of the tube. Notably, we have shown that the common approach using the the spatially dependent diffusion coefficient $D(x,f)$ fails for extreme corrugated tube geometries.

\acknowledgments
\noindent 
This article is dedicated to the memory of Dr. Frank Moss. Frank was an outstanding and enthusiastic scientist who was a true pioneer in the field of stochastic resonance \cite{Moss1985,Moss1986} and in the application of stochastic nonlinear dynamics to biological systems \cite{Moss2002,Moss2007,Moss2008}. 
We will always remember him.

This work has been supported by the VW
Foundation via project I/83903 (L.S.-G., S.M.) and I/83902 (P.H.,
G.S.). P.H. acknowledges the support the excellence cluster ''Nanosystems Initiative Munich'' (NIM).


%

\end{document}